\documentclass[preprint,prl,superscriptaddress,showpacs]{revtex4}
\usepackage{graphicx}
\usepackage{dcolumn}
\usepackage{bm}

\begin{document}


\title{Detection of high-energy gamma rays from winter thunderclouds}

\author{H. Tsuchiya}
\affiliation{Cosmic Radiation Laboratory, Riken, 2-1, Hirosawa, Wako, 
Saitama 351-0198, Japan}
\author{T. Enoto}
\affiliation{Department of Physics, University of Tokyo, 7-3-1, Hongo, 
Bunkyo-ku, Tokyo 113-0033, Japan}
\author{S. Yamada}
\affiliation{Department of Physics, University of Tokyo, 7-3-1, Hongo, 
Bunkyo-ku, Tokyo 113-0033, Japan}
\author{T. Yuasa}
\affiliation{Department of Physics, University of Tokyo, 7-3-1, Hongo, 
Bunkyo-ku, Tokyo 113-0033, Japan}
\author{M. Kawaharada}
\affiliation{Cosmic Radiation Laboratory, Riken, 2-1, Hirosawa, Wako, 
Saitama 351-0198, Japan}
\author{T. Kitaguchi}
\affiliation{Department of Physics, University of Tokyo, 7-3-1, Hongo, 
Bunkyo-ku, Tokyo 113-0033, Japan}
\author{M. Kokubun}
\affiliation{Department of High Energy Astrophysics, 
Institute of Space and Astronautical 
Science, JAXA, 3-1-1, Sagamihara, Kanagawa 229-8501, Japan}
\author{H. Kato}
\author{M. Okano}
\affiliation{Cosmic Radiation Laboratory, Riken, 2-1, Hirosawa, Wako, 
Saitama 351-0198, Japan}
\author{S. Nakamura}
\affiliation{Department of Physics, Tokyo University of Science, 1-3, Kagurazaka, 
Shinjuku-ku, Tokyo 162-8601, Japan}
\author{K. Makishima}
\affiliation{Cosmic Radiation Laboratory, Riken, 2-1, Hirosawa, Wako, 
Saitama 351-0198, Japan}
\affiliation{Department of Physics, University of Tokyo, 7-3-1, Hongo, Bunkyo-ku, 
Tokyo 113-0033, Japan}

\date{\today}

\begin{abstract}
A report is made on a comprehensive observation of a burst-like $\gamma$-ray emission from
thunderclouds on the Sea of Japan, during strong thunderstorms on 2007 January 6.
The detected emission, lasting for $\sim$40 seconds, preceded cloud-to-ground lightning discharges. 
The burst spectrum, extending to 10 MeV, can be interpreted as consisting of bremsstrahlung 
photons originating from relativistic electrons.
This ground-based observation provides first clear evidence that strong electric fields in thunderclouds 
can continuously accelerate electrons beyond 10 MeV prior to lightning discharges.  
\end{abstract}

\maketitle

\section{Introduction\label{sec:intro}}
As first suggested by \citet{wilson}, 
intense electric fields in thunderclouds and lightning are considered to accelerate 
charged particles to relativistic energies, which in turn will produce Bremsstrahlung photons. 
Thus, detailed investigation of such photons is expected to provide 
a valuable key to the particle acceleration process in strong electric fields. 

Thunderclouds are found to have electric fields reaching 100 $\mathrm{kVm^{-1}}$~\cite{ef1}, 
or sometimes even  $>$ 400 $\mathrm{kVm^{-1}}$~\cite{ef3}. 
Usually, frequent collisions in the dense atmosphere will not allow electrons to be accelerated even 
under such a strong electric field. However, seed electrons more energetic than $\sim$ 100 keV, if once produced 
by, e.g., cosmic rays, can gain energies from the electric field fast enough to overcome the collisional energy losses.
Thus, such electrons are accelerated and multiplied in the atmosphere, because their mean free paths 
get progressively longer, up to $\sim$ 50 m, as they acquire higher energies. 
This is the concept of runaway electron avalanche mechanism~\cite{RAModel}. 
When these conditions are fulfilled, electrons can be accelerated in thunderclouds to several tens MeV energies or 
higher, and then produce Bremsstrahlung $\gamma$-rays.

Actually, sudden non-thermal x-/$\gamma$-ray emissions from upper atmosphere, 
called terrestrial $\gamma$-ray flashes, have been observed at equatorial latitudes by near Earth satellites~\cite{BATSE,RHESSI}. 
The spectra of the flashes are roughly expressed by power-law functions with the photon indices of $\sim -1$~\cite{model_TGF1}, 
some of them extending up to 10 $-$ 20 MeV. Now they have been interpreted as arising from high-altitude electrical 
discharges above thunderstorms. In addition, ground-based observatories have also detected similar photon bursts from 
natural~\cite{moore,DW_exp_summber} and rocket-triggered~\cite{DW_exp_science,DW_exp_GRL2} lightning, revealing 
that the bursts are mostly associated with lightning discharge processes. 
The burst spectra in the rocket triggered lightning mostly extend up to $\sim$ 250 keV, 
or to 10 MeV~\cite{DW_exp_GRL1} on rare occasions. 
These bursts including the terrestrial $\gamma$-ray flashes have typical duration of milliseconds or less.

In addition to those short-duration bursts, more prolonged radiation from thunder activity, lasting for a few 
tens seconds to minutes, have been observed by detectors at high mountains during summer seasons~\cite{EAS,NORIKURA}. 
Also radiation-monitoring posts, arranged in nuclear power plants in the coastal area of Sea of Japan, have 
frequently detected such prolonged events associated with winter thunderstorms~\cite{MONJU}. 
However, compared with the short duration events, the longer duration ones have remained much less understood, 
due to inadequate information on, e.g., the particle species, their arrival direction and energy spectrum.
Here we show one observation revealing those essential features of the prolonged burst.
\section{Experiment\label{sec:exp}}
\begin{figure}
\includegraphics[width=0.50\textwidth]{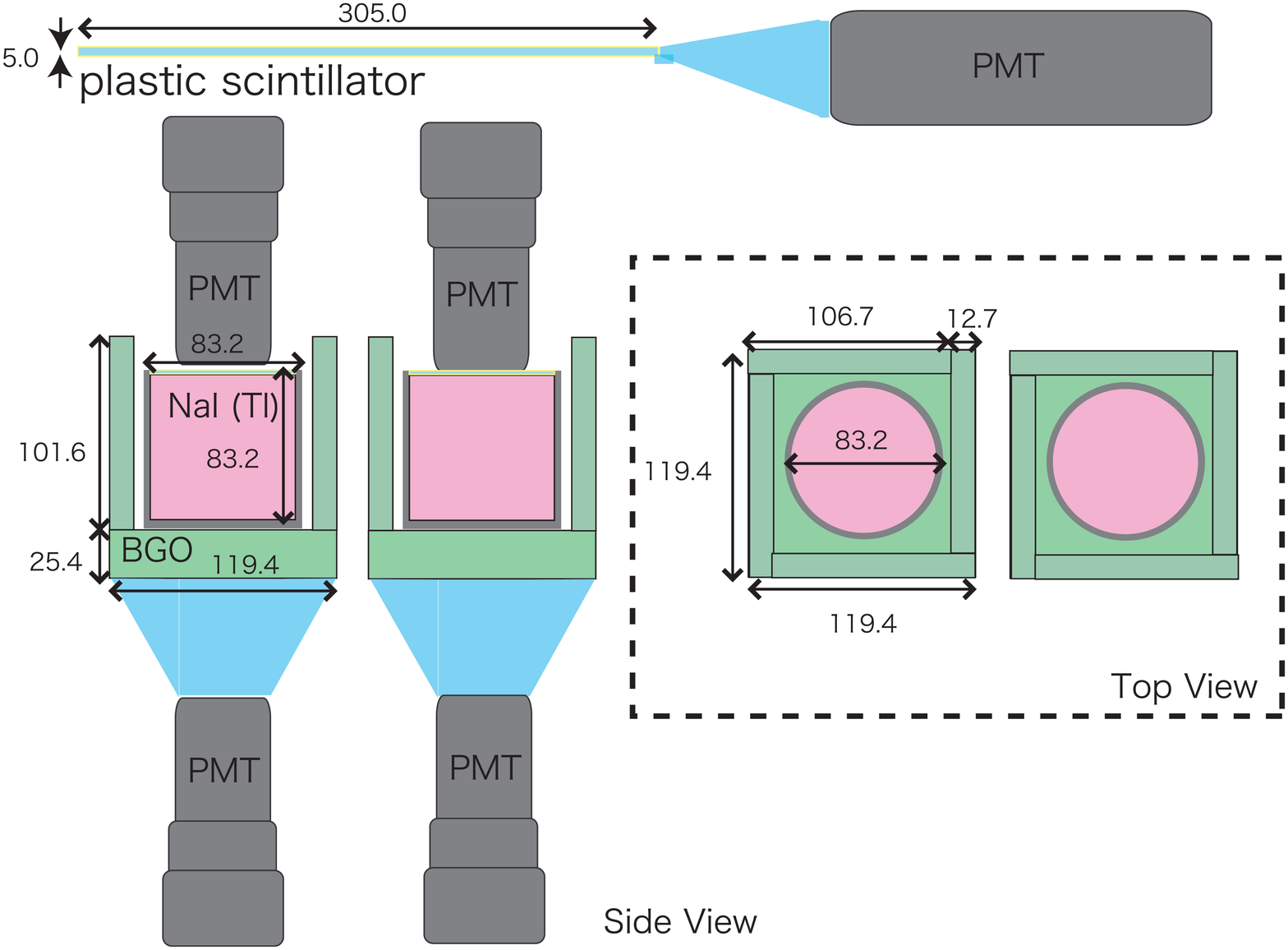}
\caption{%
Schematic views of the two main sensors and the plastic scintillator 
of Detector-A. PMT means "photomultiplier". 
Numbers represent scales in millimeter.}\label{fig:system_A}
\end{figure}

Kashiwazaki-Kariwa nuclear power plant is located in the coastal area of Sea of Japan in Niigata prefecture, 
where thunder activity is very high in November through January. 
Installed at the rooftop of a building in this power plant, our new automated radiation detection system has been continuously and 
successfully operated since 2006 December 22. The system consists of two independent and complementary parts, Detector-A and Detector-B, 
placed with a separation of  $\sim$ 10 m. 
Designed after the Hard X-ray Detector experiment~\cite{HXD} on board the 5th Japanese cosmic x-ray satellite Suzaku,
Detector-A consists of two identical main detectors, each having a structure 
shown in Fig.~\ref{fig:system_A}; it comprises a  cylindrical NaI scintillator operated in 40 keV $-$ 3 MeV, surrounded by a 
well-type BGO ($\mathrm{Bi_{4}Ge_{3}O_{12}}$) scintillator shield counter of which the average thickness is around 2.0 cm.
Placed above the two main counters is a 5 mm thick plastic scintillator. 
Detector-A records the energy deposit of each event, and its arrival time with a time resolution of 10 $\mu$s. 
As an important feature of Detector-A, the surrounding BGO shields prevent low-energy environmental radiations 
from impinging on the NaI scintillator.
Furthermore, using the signals of the BGO shields, we can roughly know where an incident particle comes from.
When the NaI scintillator is triggered by an incident particle and the NaI signals 
does not coincide with the BGO ones, the incident particle would generally arrive from the vertical direction 
neither the horizontal one nor the ground since the BGO shields firmly cover side and bottom of each NaI scintillator. 
Thus, using the BGO signals in anti-coincidence, 
we can reduce background of the NaI scintillator, and give it broad directional sensitivity towards the vertical direction.

Aiming at higher energy ranges, Detector-B utilizes NaI and CsI scintillation counters observing in 40 keV $-$ 10 MeV 
and 600 keV $-$ 80 MeV, respectively. With a spherical shape, both work as omni-directional monitors without anti-coincidence shields. 
Each of them accumulates pulse-height spectra in 1024 channels every 6 seconds, and record broad-band pulse counts every second. 

In addition to Detectors A and B, we measure environmental light and electric field.
Three light sensors are used to measure optical intensity in different directions,
each consisting of a Si photo-diode (HAMAMATSU S1226-8BK) and a handmade analog circuit.
Output signals of each light sensor are fed to a 12 bit ADC, and recorded 
as ADC values every 0.1 sec.
The electric field sensor is a commercial product (BOLTEK EFM-100), and its analog output
is collected by a 12 bit ADC every 1 sec and converted to the electric field strength
between $\pm$ 10 $\mathrm{kVm^{-1}}$ with a resolution of 5 $\mathrm{Vm^{-1}}$.
Details of our system are reported by \citet{Mth_Enoto}, and will be published elsewhere. %
\section{Results\label{sec:res}}
\begin{figure}
\includegraphics[width=0.50\textwidth]{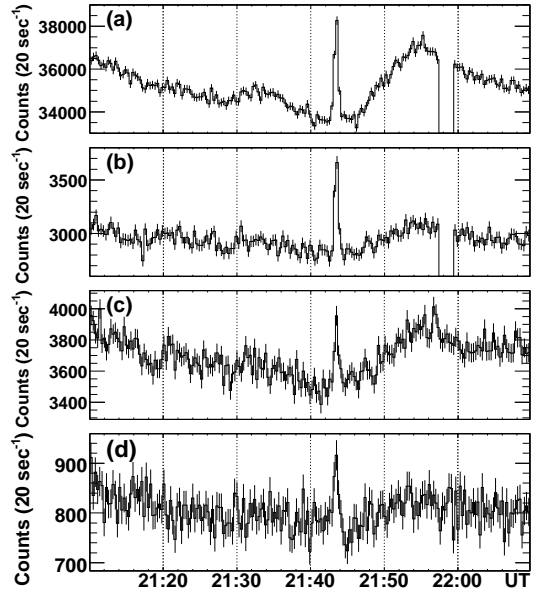}
\caption{%
Count rates per 20 seconds of 4 inorganic scintillators 
between 21:10 and 22:10 UT on 2007 January 6. All abscissa represent 
universal time, and all errors are statistical 1$\sigma$. Panels (a) 
and (b) show the count rates in $>$ 40 keV data from the BGO and NaI 
scintillators of Detector-A, respectively. Panels (c) and (d) give 
$>$ 40 keV and $>$ 600 keV data from the NaI and CsI counters of 
Detector-B, respectively. The gaps in panels (a) and (b) are due to 
regular interruption of data acquisition of Detector-A every hour.}
\label{fig:4scinti}
\end{figure}

On 2007 January 6, the coastal area of Sea of Japan was covered with one of the strongest thunderstorms in this winter. 
Figure~\ref{fig:4scinti} shows count rate histories of the BGO and NaI scintillators of Detector-A, and of the NaI and 
CsI counters of Detector-B, obtained from 21:10 to 22:10 UT on this day (6:10 to 7:10 local standard time on January 7). 
Superposed on gradual background changes due to 
fall-outs of radioactive Radon, we have detected a remarkable count enhancement at about 21:43, lasting for $\sim$ 1 min.  
This event, or "burst", is thus highly significant in all the inorganic scintillators of our system.

\begin{figure}
\includegraphics[width=0.50\textwidth]{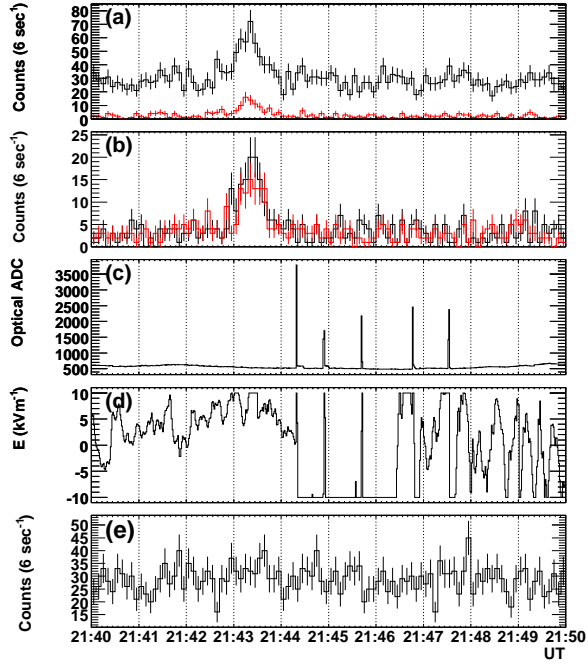}
\caption{%
Detailed count rate histories of the radiation sensors, and the light and 
electric field monitors, between 21:40 and 21:50 UT. All abscissa are 
universal time. All bins except those in panels (c) and (d) have $1\sigma$ 
statistical errors. (a) Count histories per 6 seconds in $>$ 3 MeV energies 
from the main NaI sensors of Detector-A, without (black) and with (red) the 
BGO anticoincidence. (b) 3 $-$ 10 MeV counts every 6 seconds from the 
NaI (black) and CsI (red) scintillators of Detector-B. (c) One-second optical 
data variations. (d) One-second electric field data variations. (e) $>$ 1 MeV 
counts every 6 seconds from the plastic scintillator of Detector-A.}
\label{fig:detail_lc}
\end{figure}

Figure~\ref{fig:detail_lc} gives details of the count-rate histories between 21:40 and 21:50 UT. Records of light and 
electric fields are also shown. The signal in Detector-A is most prominent in the $>$ 3 MeV range covered by the last 
overflow bin [Fig .~\ref{fig:detail_lc}(a)], where the excess counts in a 36 sec of 21:43:09 $-$ 21:43:45 UT,
calculated by subtracting the background from the total counts during the burst,
become $129 \pm 17$ and $43.6 \pm 3.5$, without and with the BGO anticoincidence, respectively. The NaI and CsI counters 
of Detector-B recorded 3 $-$ 10 MeV excess counts of $74.2 \pm 9.6$ 
and $53.3 \pm 8.4$, respectively [Fig.~\ref{fig:detail_lc}(b)]. Therefore, the burst intensity from the two Detectors 
(without anticoincidence for A) agrees with one another, since the values from Detector-A sum the two counters. 

At 21:44:18 UT, or 70 seconds after the burst onset, the optical data rapidly increased for about 1 second 
[Fig.~\ref{fig:detail_lc}(c)], with a rise time less than the sampling time which is 100 ms. In coincidence with this 
flash, the electric filed quickly changed its polarity from positive values ($\ge 5$ $\mathrm{kVm^{-1}}$) during the burst
to negative ($\le - 10$ $\mathrm{kVm^{-1}}$), although the signal is saturated at $\pm 10$ $\mathrm{kVm^{-1}}$ 
due to the limited range of the electric field sensor. 
Thus, the first flash can be interpreted 
as a lightning discharge; so are the subsequent four flashes. The burst had already ceased by the first 
lightning, and no increase accompanied any of the total five discharges. Thus, the observed event appears to be 
associated with thunderclouds themselves, rather than to the lightning discharges.

During the burst, the plastic scintillator of Detector-A [Fig.~\ref{fig:detail_lc}(e)], with its lower threshold set at 1 MeV, 
showed no significant excess signals. Quantitatively, the number of events 
detected by the plastic scintillator with an area of 464 $\mathrm{cm^{2}}$ in the 36 sec is $185 \pm 14$, which 
is consistent with the number, 191, predicted by its longer-term average count. 
The lack of excess signals in the plastic scintillator, 
together with the long-burst duration, excludes the burst being due to electrical noise.
Furthermore, we can conclude that the burst is dominated by $\gamma$-rays, since the thin plastic scintillator has high 
detection efficiency for charged particles but is nearly transparent to $\gamma$-rays. 
Another important information, derived from the data, is that the photons 
came from a sky, probably covered with thunderclouds, rather than from the ground. Actually, excess count ratio between 
NaI~[Fig.\ref{fig:4scinti}(b)] and BGO~[Fig.\ref{fig:4scinti}(a)] of Detector-A, $\sim 0.18 \pm 0.03$, is significantly higher 
than their normal count ratio, 0.08, which is due mostly to environmental radioactivity.

\begin{figure*}[tbh]
\includegraphics[width=0.80\textwidth]{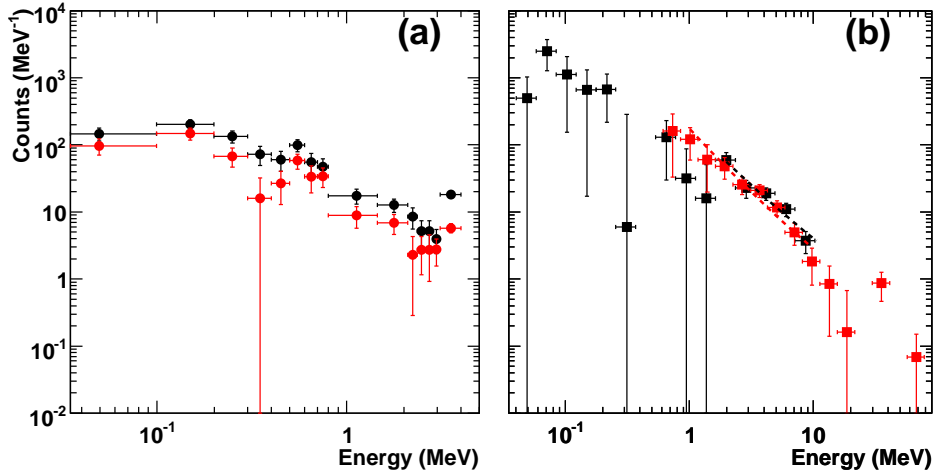}
\caption{%
Background-subtracted photon energy spectra accumulated over the 36 sec. 
All horizontal and vertical axes show the photon energy and the counts 
per each energy bin, respectively. Error bars are statistical $1\sigma$.
(a) Spectra from Detector-A. Black and red circles show data of the main 
NaI counters (summed over the two units), without and with the BGO 
anticoincidence. (b) Spectra from Detector-B. Black and red squares are 
from NaI and CsI counters, respectively. Black and red dashed lines show 
predictions of the best-fit incident power-law model, determined over the 
energy range where the dashed lines are drawn.}
\label{fig:spectra}
\end{figure*}

Figure~\ref{fig:spectra} shows pulse-height spectra of the burst from Detector-A and B, accumulated for 
the same 36 sec; we subtracted background averaged over 21:30 $-$ 21:40 UT and 21:45 $-$ 21:55 UT. Thus, the excess 
counts in both systems exhibit very hard continuum spectra. The spectra of Detector-B, in particular, extend up 
to 10 MeV. Supposing that the $\gamma$-rays came from the thunderclouds, these spectra imply that electrons were accelerated 
in the thunderclouds at least to 10 MeV, or to higher energies in order to efficiently produce the 10 MeV photons.

Based on results from an experiment~\cite{DW_exp_GRL1}, and from a Monte Carlo simulation~\cite{Torii_cal},
we may assume that the photon number spectrum arriving at the top of our system has a power-law form 
as $\alpha \times E^{\beta}$, where $\alpha$ is normalization in units 
of photons $\mathrm{MeV^{-1}cm^{-2}}$, $\beta$ is the photon index, and $E$ is the photon energy in MeV. In this work, 
the data points in 600 keV $-$ 10 MeV where both NaI and CsI counters can cover are selected [Fig.~\ref{fig:spectra}(b)]. 
Then, convolving this model with the detector responses, and fitting it to the observed spectra (but excluding data points with 
too large errors), we obtained $\alpha = 2.63 \pm 0.92$ and $\beta = -1.58 \pm 0.23$ ($\chi^{2}/d.o.f. = 3.3/3$) 
from NaI of Detector-B, while $\alpha = 1.93 \pm 0.42$  and $\beta = -1.69 \pm 0.15$ ($\chi^{2}/d.o.f. = 5.3/6$) 
from CsI. Thus, the two sets of results are consistent. 
Weighted means of the two estimates become 
$\alpha = 2.04 \pm 0.38$ and $\beta = -1.66 \pm 0.13$. The spectra from Detector-A are generally consistent with this.
In reality, the observed photons must be produced at some distance, and propagate 
in the atmosphere while they are attenuated mainly via Compton scattering. When corrected for the mild energy 
dependence of the Compton cross section, 
the photon index softens to $-2.01 \pm 0.14$, $-2.44 \pm 0.15$, and $-3.03 \pm 0.16$, 
assuming the source distance of 300, 600, and 1,000 m, respectively. 
\section{Discussions\label{sec:dis}}
With respect to the duration and its precedence to the lightning, 
the present event is similar to one event observed by \citet{MONJU}, 
on another nuclear plant located on the same coastal area with a 
separation of 300 km. However, compared with their results, the present 
work provides a few new findings. That is, we have confirmed that 
the signal is dominated by $\gamma$-rays arriving from the sky direction. 
In addition, our results reveal that the photon spectrum extends up to 10 MeV.

The photon spectrum detected in the present work, extending up to 10 MeV, 
clearly indicates a non-thermal emission mechanism. The most promising 
candidate process is Bremsstrahlung. Then, we can draw a following possible 
scenario to explain the present observation~\cite{Mth_Enoto}. Generally, 
a winter thundercloud along the Sea of Japan is considered to have 
3-layer (positive-negative-positive) charge structure in the vertical 
direction~\cite{3struct}. Then, the lower part of such thunderclouds is 
positively charged. Electrons, first produced by cosmic rays, will be 
accelerated by the runaway electron avalanche process~\cite{RAModel} from 
the middle part to the cloud bottom (or top), emitting Bremsstrahlung 
$\gamma$-rays toward the ground (or sky). Since the $\gamma$-rays must be 
relativistically beamed into a narrow cone ($\sim3^\circ$ half-cone angle 
for a 10 MeV electron), we can detect such photons only under rare conditions 
when the cone happens to fall on our detector.

Let us briefly evaluate energetics of the above scenario. Since we 
observed a fluence of $1 - 10$ MeV photons as $n_\mathrm{p} = 2.4\times 10^{4}$ $\mathrm{m^{-2}}$, 
the total number of these photons at the source $N_\mathrm{p}$ becomes
$d^2 n_\mathrm{p} \epsilon^{-1} \Delta \Omega$, where $d$ is the source distance,
$\epsilon$ is the atmospheric attenuation factor,
and $\Delta \Omega$ is a solid angle of the possibly beamed emission.
The values of $\epsilon$ for 3 MeV photons are calculated, e.g., 
as 0.3, 0.08, and 0.01 for $d=300$, 600, and 1000 m, respectively.
Employing $\Delta \Omega = 8.6\times 10^{-3}$
(or equivalently a half-cone angle of $\sim 3^\circ$)
which is appropriate for 10 MeV electrons,
we obtain $N_\mathrm{p} \sim 10^8 (d/300\mathrm{m})^{2}(0.3/\epsilon)$.
Since a 10 MeV electron traveling $\sim$ 50 m in the atmosphere
is expected to lose a minor fraction, $\eta \sim 10^{-2}$~\cite{Rossi},
of its energy into 3 MeV bremsstrahlung photons, the overall energy 
of the produced relativistic electrons is estimated as
\begin{equation}
E_\mathrm{e} = N_\mathrm{p}/\eta \times 10~\mathrm{MeV} 
\sim 10^{-2} (d/300\mathrm{m})^{2}(0.3/\epsilon)~~\mathrm{J.} 
\end{equation}
Of course, the estimate is subject to rather large uncertainties,
because we may not have observed the peak of the radiation beam.
Nevertheless, the estimated $E_\mathrm{e}$ is a negligible fraction
of an energy associated with a single lightning, $10^{7} - 10^{10}$ J~\cite{E_l},
ensuring that the scenario is energetically feasible.

In summary, we have successfully shown that electrons are accelerated 
beyond 10 MeV in thunderclouds, for an extended period, prior to lightning 
discharges rather than in coincidence with them. Thus, this kind of observation 
of non-thermal x-/$\gamma$-ray emissions associated with thunderclouds and 
lightning will provide a valuable key to the particle acceleration processes 
in strong electric fields.
\begin{acknowledgments}
T. Torii kindly gave us helpful advice on our experiment. 
We greatly thank members of radiation safety group of Kashiwazaki-Kariwa 
power station, Tokyo Electric Power Company, including K. Oshimi, 
T. Nakamura, K. Inomata, and Y. Iwasaki, for support of our experiment.  
We are grateful to S. Otsuka and Y. Ikegami for design and construction of 
housings of our system. This work is supported in part by the Special Research 
Project for Basic Science in RIKEN, titled ``Investigation of Spontaneously 
Evolving Systems''. The work is also supported in part by Grant-in-Aid for 
Scientific Research (S), No.18104004. 
\end{acknowledgments}
\newpage 

%
%
\clearpage
\clearpage
%
%
\clearpage
%
%
\clearpage
%
%


\begin{thebibliography}{20}
\expandafter\ifx\csname natexlab\endcsname\relax\def\natexlab#1{#1}\fi
\expandafter\ifx\csname bibnamefont\endcsname\relax
  \def\bibnamefont#1{#1}\fi
\expandafter\ifx\csname bibfnamefont\endcsname\relax
  \def\bibfnamefont#1{#1}\fi
\expandafter\ifx\csname citenamefont\endcsname\relax
  \def\citenamefont#1{#1}\fi
\expandafter\ifx\csname url\endcsname\relax
  \def\url#1{\texttt{#1}}\fi
\expandafter\ifx\csname urlprefix\endcsname\relax\def\urlprefix{URL }\fi
\providecommand{\bibinfo}[2]{#2}
\providecommand{\eprint}[2][]{\url{#2}}
%
\bibitem[{\citenamefont{Wilson}(1925)}]{wilson}
  \bibinfo{author}{\bibfnamefont{C.~T.~R.} \bibnamefont{Wilson}},
  \bibinfo{journal}{Proc. of Phys. Soc. London} \textbf{\bibinfo{volume}{37}},
  \bibinfo{pages}{32} (\bibinfo{year}{1925}).
%
\bibitem[{\citenamefont{Marshall et~al.}(1995)\citenamefont{Marshall, Rust, and Stolzenburg}}]{ef1}
  \bibinfo{author}{\bibfnamefont{T.~C.} \bibnamefont{Marshall}},
  \bibinfo{author}{\bibfnamefont{W.~D.} \bibnamefont{Rust}}, \bibnamefont{and}
  \bibinfo{author}{\bibfnamefont{M.}~\bibnamefont{Stolzenburg}},
  \bibinfo{journal}{J. Geophys. Res.} \textbf{\bibinfo{volume}{100}},
  \bibinfo{pages}{1001} (\bibinfo{year}{1995}).
%
\bibitem[{\citenamefont{Marshall et~al.}(2005)}]{ef3}
  \bibinfo{author}{\bibfnamefont{T.~C.} \bibnamefont{Marshall} {\it et al.}}, 
  \bibinfo{journal}{Geophys. Res. Lett.} 
  \textbf{\bibinfo{volume}{32}}, \bibinfo{pages}{L03813} (\bibinfo{year}{2005}).
%
\bibitem[{\citenamefont{Gurevich et~al.}(1992)\citenamefont{Gurevich, Milikh, and Roussel-Dupre}}]{RAModel}
  \bibinfo{author}{\bibfnamefont{A.~V.} \bibnamefont{Gurevich}},
  \bibinfo{author}{\bibfnamefont{G.~M.} \bibnamefont{Milikh}},
  \bibnamefont{and}
  \bibinfo{author}{\bibfnamefont{R.}~\bibnamefont{Roussel-Dupre}},
  \bibinfo{journal}{Phys. Lett. A} \textbf{\bibinfo{volume}{165}},
  \bibinfo{pages}{463} (\bibinfo{year}{1992}).
%
\bibitem[{\citenamefont{Fishman et~al.}(1994)}]{BATSE}
  \bibinfo{author}{\bibfnamefont{G.~J.} \bibnamefont{Fishman} {\it et al.}},
  \bibinfo{journal}{Science}
  \textbf{\bibinfo{volume}{264}}, \bibinfo{pages}{1313} (\bibinfo{year}{1994}).
\bibitem[{\citenamefont{Smith et~al.}(2005)}]{RHESSI}
  \bibinfo{author}{\bibfnamefont{D.~M.} \bibnamefont{Smith} {\it et al.}},
  \bibinfo{journal}{Science} \textbf{\bibinfo{volume}{307}},
  \bibinfo{pages}{1085} (\bibinfo{year}{2005}).
%
\bibitem[{\citenamefont{Nemiroff et~al.}(1997)\citenamefont{Nemiroff, Bonnell, and Norris}}]{model_TGF1}
  \bibinfo{author}{\bibfnamefont{R.~J.} \bibnamefont{Nemiroff}},
  \bibinfo{author}{\bibfnamefont{J.~T.} \bibnamefont{Bonnell}}, \bibnamefont{and} 
  \bibinfo{author}{\bibfnamefont{J.~P.} \bibnamefont{Norris}}, 
  \bibinfo{journal}{J. Geophys. Res.}
  \textbf{\bibinfo{volume}{102}}, \bibinfo{pages}{9659} (\bibinfo{year}{1997}).
%
\bibitem[{\citenamefont{Moore et~al.}(2001)}]{moore}
  \bibinfo{author}{\bibfnamefont{C.~B.} \bibnamefont{Moore} {\it et al.}},
  \bibinfo{journal}{Geophys. Res. Lett.} \textbf{\bibinfo{volume}{28}},
  \bibinfo{pages}{2141} (\bibinfo{year}{2001}).
%
\bibitem[{\citenamefont{Dwyer et~al.}(2005)}]{DW_exp_summber}
\bibinfo{author}{\bibfnamefont{J.~R.} \bibnamefont{Dwyer} {\it et al.}},
\bibinfo{journal}{Geophys. Res. Lett.} \textbf{\bibinfo{volume}{32}},
\bibinfo{pages}{L01803} (\bibinfo{year}{2005}).
%
\bibitem[{\citenamefont{Dwyer et~al.}(2003)}]{DW_exp_science}
\bibinfo{author}{\bibfnamefont{J.~R.} \bibnamefont{Dwyer} {\it et al.}},
\bibinfo{journal}{Science}  \textbf{\bibinfo{volume}{299}}, \bibinfo{pages}{694} (\bibinfo{year}{2003}).
%
\bibitem[{\citenamefont{Dwyer et~al.}(2004)}]{DW_exp_GRL2}
\bibinfo{author}{\bibfnamefont{J.~R.} \bibnamefont{Dwyer} {\it et al.}},
 \bibinfo{journal}{Geophys. Res. Lett.} \textbf{\bibinfo{volume}{31}}, \bibinfo{pages}{L05118}
  (\bibinfo{year}{2004}).
%
\bibitem[{\citenamefont{Dwyer et~al.}(2004)}]{DW_exp_GRL1}
\bibinfo{author}{\bibfnamefont{J.~R.} \bibnamefont{Dwyer} {\it et al.}},
 \bibinfo{journal}{Geophys. Res. Lett.} \textbf{\bibinfo{volume}{31}}, \bibinfo{pages}{L05119}
  (\bibinfo{year}{2004}).
%
\bibitem[{\citenamefont{Chubenko et~al.}(2000)}]{EAS}
\bibinfo{author}{\bibfnamefont{A.~P.} \bibnamefont{Chubenko} {\it et al.}},
  \bibinfo{journal}{Phys. \ Lett. \ A}
  \textbf{\bibinfo{volume}{275}}, \bibinfo{pages}{90} (\bibinfo{year}{2000}).
%
\bibitem[{\citenamefont{Muraki et~al.}(2004)}]{NORIKURA}
\bibinfo{author}{\bibfnamefont{Y.}~\bibnamefont{Muraki} {\it et al.}},
  \bibinfo{journal}{Phys. Rev. D}\textbf{\bibinfo{volume}{69}}, \bibinfo{pages}{123010}
  (\bibinfo{year}{2004}).
%
\bibitem[{\citenamefont{Torii et~al.}(2002)\citenamefont{Torii, Takeishi, and Hosono}}]{MONJU}
\bibinfo{author}{\bibfnamefont{T.}~\bibnamefont{Torii}},
  \bibinfo{author}{\bibfnamefont{M.}~\bibnamefont{Takeishi}}, \bibnamefont{and}
  \bibinfo{author}{\bibfnamefont{T.}~\bibnamefont{Hosono}},
  \bibinfo{journal}{J. Geophys. Res.} \textbf{\bibinfo{volume}{107}},
  \bibinfo{pages}{4324} (\bibinfo{year}{2002}).
%
\bibitem[{\citenamefont{Takahashi et~al.}(2007)}]{HXD}
  \bibinfo{author}{\bibfnamefont{T.}~\bibnamefont{Takahashi} {\it et al.}},
  \bibinfo{journal}{Publ. Astron. Soc. Japan} \textbf{\bibinfo{volume}{59}},
  \bibinfo{pages}{S35} (\bibinfo{year}{2007}).
%
\bibitem[{\citenamefont{Enoto}(2007)}]{Mth_Enoto}
  \bibinfo{author}{\bibfnamefont{T.}~\bibnamefont{Enoto}},
  \bibinfo{journal}{Master Thesis, University of Tokyo (in Japanese)} (\bibinfo{year}{2007}).
%
\bibitem[{\citenamefont{Torii et~al.}(2004)}]{Torii_cal}
  \bibinfo{author}{\bibfnamefont{T.}~\bibnamefont{Torii} {\it et al.}},
  \bibinfo{journal}{Geophys. Res. Lett.} \textbf{\bibinfo{volume}{31}},
  \bibinfo{pages}{L05113} (\bibinfo{year}{2004}).
%
\bibitem[{\citenamefont{Kitagama and Michimoto}(1994)\citenamefont{Kitagama and Michimoto}}]{3struct}
  \bibinfo{author}{\bibfnamefont{N.}~\bibnamefont{Kitagawa}} \bibnamefont{and}
  \bibinfo{author}{\bibfnamefont{K.}~\bibnamefont{Michimoto}},
  \bibinfo{journal}{J. Geophys. Res.} \textbf{\bibinfo{volume}{99}},
  \bibinfo{pages}{10713} (\bibinfo{year}{1994}).
%
\bibitem[{\citenamefont{Rossi}(1965)}]{Rossi}
\bibinfo{author}{\bibfnamefont{B.}~\bibnamefont{Rossi}},
  \emph{\bibinfo{title}{High-Energy Particles}}
  (\bibinfo{publisher}{Prentice-Hall Inc., Englewood Cliffs, N.J.},
  \bibinfo{year}{1965}), p. \bibinfo{pages}{51}.
%
\bibitem[{\citenamefont{Marshall and Stolzenburg}(2001)\citenamefont{Marshall and Stolzenburg}}]{E_l}
  \bibinfo{author}{\bibfnamefont{T.~C.} \bibnamefont{Marshall}} \bibnamefont{and}
  \bibinfo{author}{\bibfnamefont{M.}~\bibnamefont{Stolzenburg}},
  \bibinfo{journal}{J. Geophys. Res.} \textbf{\bibinfo{volume}{106}},
  \bibinfo{pages}{4757} (\bibinfo{year}{2001}).
%
\end{thebibliography}
\end{document}